\documentclass[10pt,aps,two column,ajp]{revtex4}   

\usepackage{amsmath}    
\usepackage{amsfonts}  
\usepackage{amssymb}
\usepackage{dsfont}
\usepackage{mathrsfs}
\usepackage{graphicx}   
\usepackage{bigints}

\begin{document}

\newcommand{\vett}[1]{\mathbf{#1}}
\newcommand{\uvett}[1]{\hat{\vett{#1}}}
\newcommand{\beq}{\begin{equation}}
\newcommand{\eeq}{\end{equation}}
\newcommand{\barr}{\begin{eqnarray}}
\newcommand{\earr}{\end{eqnarray}}

\title{The Faddeev-Popov Method Demystified}

\author{Marco Ornigotti}
\affiliation{Institute of Applied Physics, Friedrich-Schiller University, Jena, Max-Wien Platz 1, 07743 Jena, Germany} 
 \email{marco.ornigotti@uni-jena.de}   
 
\author{Andrea Aiello}
\affiliation{Max Planck Institute for the Science of Light, G$\ddot{u}$nther-Scharowsky-Strasse 1/Bau24, 91058 Erlangen,
Germany} 
\affiliation{Institute for Optics, Information and Photonics, University of Erlangen-Nuernberg, Staudtstrasse 7/B2, 91058 Erlangen, Germany}

\begin{abstract}
We discuss how to implement the Legendre transform using the Faddeev-Popov method of Quantum Field Theory. By doing this, we provide an alternative way to understand the essence of the Faddeev-Popov method, using only concepts that are very familiar to the students (such as the Legendre transform), and without needing any reference to Quantum Field Theory. Two examples of Legendre transform calculated with the Faddeev-Popov method are given to better clarify the point.
\end{abstract}

\date{\today}

\maketitle

\section{introduction}
Students learn in the early days of their studies that the electromagnetic field can be described by a single quantity,  the vector potential $A_{\mu}$, instead of the separate electric and magnetic fields \cite{jackson}. However, two distinct vector potentials $A_{\mu}$ and $A_{\mu}'=A_{\mu}+\partial_{\mu}\Lambda(x)$ (where $\Lambda(x)$ is a scalar field, the gauge field, that satisfies the Helmholtz equation) generate the same electric and magnetic fields. In order to remove this ambiguity, one should first \emph{fix the gauge}, namely determine uniquely the gauge function $\Lambda(x)$. Once the gauge has been chosen, the correspondence between the vector potential and the electromagnetic fields is made unique. The electromagnetic field, however, is not the only gauge field in Nature. In fact, all the fundamental forces, such as the weak force (responsible for radioactive decay and nuclear fusion of subatomic particles)  and the strong force (that ensures the stability of ordinary matter) are both described by gauge fields \cite{weinberg}.


Since the quantum theory of a gauge field is often described by means of the path integral method \cite{das}, a generalization of the concept of gauge fixing within this framework is of paramount importance for the development of the theory itself. The Faddeev-Popov method exactly complies with this needing. From a rigorous point of view, this method consists in using a gauge fixing condition to reduce the number of allowed orbits of the mathematical configurations that represent a given physical system to a smaller set, where all the orbits are related by a smaller gauge group symmetry \cite{weinberg}. In simpler words, the Faddeev-Popov method consists in applying a constraint to the considered field, that automatically implements the gauge fixing condition, thus determining the field unambiguously \cite{brown}.

Usually, this method is object of doubts and of difficult understanding for the students (either graduate or undergraduate) that encounter it for the first time, and the development of this method reported in standard quantum field theory books \cite{brown,ryder,peskin} is not helpful for having a better understanding of it. In fact, most of the times it appears more like a magic trick rather than motivated by some physical argument.

The Legendre transform, on the other hand, is a very simple mathematical instrument that a student knows very well, since it finds application in a wide variety of physical problems, like the transformation between the Hamiltonian and the Lagrangian function of a system in classical mechanics \cite{goldstein}, the analysis of equilibrium states in statistical physics \cite{landau} or  the transformation between the thermodynamical functions (entropy, enthalpy, Gibbs free energy, etc.) \cite{thermo}. This mathematical tool has the advantage, with respect to the Faddeev-Popov method, to be of immediate understanding, because it involves a very well known procedure that any student has seen at least one time during his studies. The possibility of illustrating the results of the Faddeev-Popov ``magic" method in terms of a Legendre transform would be of great didactical insight, since it will allow a better understanding of the physical foundations of a not so easy to understand common method in quantum field theory.

It is then the aim of this paper to unravel this connection, by using a simple example in which the Faddeev-Popov method is used to implement the Legendre transform of a given function $f(x)$. This is possible because these two methods, apparently very different and far away from each other, share in reality the same essence, namely they both consist in applying constraint to certain physical systems: the Faddeev-Popov method fixes the field gauge by imposing the gauge condition as a constraint to the field, and the Legendre transform implements the transformation between a function $f(x)$ and its transformed pair $g(p)$  by imposing a certain constraint to the function $F(x,p)=xp-f(x)$ \cite{convex}. 

This work is organized as follows: in Sect. II  we review very briefly the Faddeev-Popov method, in the simplest case of the electromagnetic field. This allows us to write Eq. \eqref{due}, that constitutes central point for our work. In Sect. III,  a brief recall on the definition of Legendre transform is given. Finally, in Sect. IV we derive the Legendre transform formula by making use of the Faddeev-Popov rule. Conclusions are then drawn in Sect. V.
 
\section{The Faddeev-Popov method in a nutshell}
 One of the possible ways to quantize a field is the so-called path integral quantization. In this scheme, the transition probability for a physical system to evolve from an initial configuration $\phi(x_i)$ to a final configuration $\phi(x_f)$ is obtained by summing all the probability amplitudes corresponding to all possible paths in space-time that the system will take to reach the final point $x_f$ starting from the initial point $x_i$ \cite{feynman}. Then, all the transition amplitudes (\emph{n}-point functions) can be obtained from a generating functional (the path integral) of the form
\begin{equation}\label{uno}
 Z[J]=\int\mathcal{D}A_{\mu} e^{i\int d^4x\;\left(\mathcal{L}+J^{\mu}A_{\mu}\right)},
 \end{equation}
where $\mathcal{L}\equiv\mathcal{L}[A_{\mu},\partial_{\nu}A_{\mu},x]$ is the Lagrangian density,  $A_{\mu}$ represent the field variables and $J^{\mu}$ is an external current. For the free electromagnetic field the Lagrangian density is given by $\mathcal{L}=-(1/4)F_{\mu\nu}F^{\mu\nu}$, and since there are no field sources, $J^{\mu}$ has to be set to zero at the end of the calculation \cite{noteJ}. 

Following Sect. 7.2 of Ref. \cite{ryder}, we can introduce heuristically the Faddeev-Popov method by making the following observations. First of all, the integral measure $\mathcal{D}A_{\mu}=\Pi_xdA_{\mu}(x)$ in Eq. \eqref{uno} accounts for every possible $A_{\mu}$, thus including those that are connected via a gauge transformation. We can make this fact explicit by introducing the following notation:
\beq
A_{\mu}\;\rightarrow \bar{A}_{\mu}^{\Lambda}.
\eeq
This means that now we are considering  $A_{\mu}$ as a representative of the class of vector potentials that can be obtained from a given $\bar{A}_{\mu}$ by performing a gauge transformation of the type
\beq\label{gaugeA}
A_{\mu}=\bar{A}_{\mu}+\partial_{\mu}\Lambda(x).
\eeq
With this trick, we can rewrite the generating functional $Z[J]$ as follows, by separating the contributions of $\bar{A}_{\mu}$ and $\Lambda(x)$:
\beq\label{nonReg}
Z[J]=\int\mathcal{D}\bar{A}_{\mu}e^{i\int d^4x\;\left(\mathcal{L}+J^{\mu}\bar{A}_{\mu}\right)}\int\mathcal{D}\Lambda.
\eeq
The presence of the second integral $\int\mathcal{D}\Lambda$ is the one that causes $Z[J]$ to diverge, since there are infinitely many gauge fields $\Lambda(x)$ that satisfy Eq. \eqref{gaugeA}. The regularization of the integral \eqref{nonReg} is the scope of the Faddeev-Popov method. 

In order to fully understand the essence of this method, let us consider the following resolution of the identity \cite{ryder}:
\begin{equation}\label{identity}
I=\mathrm{det}\Bigg(\frac{\delta G[A^{\Lambda}]}{\delta\Lambda}\Bigg)\Bigg|_{\Lambda=0}\int\mathcal{D}\Lambda\delta(G[A^{\Lambda}]),
\end{equation}
where the derivative in round brackets has to be intended in the sense of functional derivative \cite{brown}. The Dirac delta function inside the integral is nothing else but the gauge condition that one needs to impose to the Lagrangian of the free electromagnetic field \cite{comment} in order to fix the gauge, namely

\begin{equation}\label{gaugeCond}
G[A^{\Lambda}]=\partial^{\mu}A_{\mu}+\frac{1}{e}\partial^{\mu}\partial_{\mu}\Lambda(x),
\end{equation}
where $e$ is the electron charge. Eq. \eqref{identity} is simply a generalization to the domain of functional analysis of the well known expression \cite{matBook}
\beq
1=\left[\left(\frac{\partial g(x)}{\partial x}\right)\Bigg|_{x=x_0}\right]^{-1}\int\;dx\;\delta\left(g(x)\right),
\eeq
where $x_0$ is the point at which $g(x)$ vanishes. Inserting the expansion of the unity as given by Eq. \eqref{identity} into Eq. \eqref{uno} (and dropping the bar symbol from $\bar{A}_{\mu}$) gives:
\begin{equation}\label{due}
Z[J]=\int\mathcal{D}\Lambda\int\mathcal{D}A_{\mu}\Delta[A]\delta(G[A^{\Lambda}])e^{i\int\;d^4x\left(\mathcal{L}+J^{\mu}A_{\mu}\right)},
\end{equation}
where
\beq\label{deltaD}
\Delta[A]=\mathrm{det}\Bigg(\frac{\delta G[A^{\Lambda}]}{\delta\Lambda}\Bigg)\Bigg|_{\Lambda=0}.
\eeq
This result should be compared with Eq. \eqref{nonReg}. In this case, in fact, the generating functional $Z[J]$ is regularized, thanks to the presence of the Dirac delta function, whose role is to uniquely determine the vector potential $A_{\mu}$ by suitably fixing the gauge \eqref{gaugeCond} that is given as the argument of the Dirac delta function itself. This is the essence of the Faddeev-Popov method: by inserting in the generating functional \eqref{uno} the resolution of the identity as given by Eq. \eqref{identity}, one is now able to introduce a constraint (through a Dirac delta function) that automatically fixes the gauge, thus eliminating the arbitrariness of the vector potential $A_{\mu}$ and removing the problem of overcounting the field configurations.

Let us now put this result in a more appealing form, that will be helpful for the results of the next section. To this aim, let us rewrite Eq. \eqref{due} in the following way:
\beq\label{FPalt}
Z[J]=\int\mathcal{D}\Lambda\;\mathcal{F}[A_{\mu},\Lambda;J]\Delta[A_{\mu}]\delta(G[A^{\Lambda}]),
\eeq
where
\beq
\mathcal{F}[A_{\mu},\Lambda;J]=\int\mathcal{D}A_{\mu}e^{i\int\;d^4x\left(\mathcal{L}+J^{\mu}A_{\mu}\right)},
\eeq
and the dependence of $\mathcal{F}$ on $\Lambda$ is implicitly contained in $A_{\mu}$ through Eq. \eqref{gaugeA}.  Although at a first glance Eq. \eqref{FPalt} may appear cumbersome and of little use, its form has a great advantage: it may be directly used to calculate the Legendre transform of an 
%
%
 arbitrary function $f(x)$, as we will discuss in the next section. Moreover, written in this form, the physical idea behind the Faddeev-Popov method appear more clear: given a functional $\mathcal{F}[A_{\mu},\Lambda;J]$ that depends on the particular choice of the gauge function $\Lambda(x)$, the Faddeev-Popov method consists in inserting the gauge fixing condition $G[A^{\Lambda}]$ in the form of an integral over the gauge configurations $\Lambda$. The presence of the term $\Delta[A_{\mu},0]$ simply accounts for the correct normalization term to insert in order not to modify the physics of the original system.
\section{The Legendre transform}
Before calculating the Legendre transform by means of the Faddeev-Popov method, let us first briefly recall the definition of Legendre transform and clarify its use with a simple example.
Following the standard textbook definition \cite{convex}, let us assume to have a function $f(x)$, and let us define, starting from $f(x)$, the two functions $F(x,p)=px-f(x)$ and $G(x,p)=\partial F(x,p)/\partial x$. It is then possible to define the Legendre transform $g(p)$ of $f(x)$ as:
\begin{equation}\label{leg1}
g(p)=F(x,p)|_{G(x,p)=0},
\end{equation}
where the constraint $G(x,p)=0$ is intended to be solved with respect to $x$. The reader should also keep in mind that, in order for the Legendre transform to make sense, the initial function $f(x)$ has to be convex, i.e., the function must have positive second derivative. It is also worth to be noted, that the definition of Legendre transform given here is only one of its possible form. Frequently, the Legendre transform is also defined as a constrained maximization of the function $f(x)$ \cite{convex}. For the purposes of this paper, however, this definition suites more the needings. 

As an example of calculation of the Legendre transform, let us take the simple function $f(x)=x^2/2$. We then have that $F(x,p)=px-x^2/2$, $G(x,p)=p-x$, and the function $G(x,p)$ vanishes for $x_0=p$. Substituting this into Eq. \eqref{leg1} we have the following:
\barr
g(p)&=&F(x,p)|_{G(x,p)=0}=\left(xp-\frac{x^2}{2}\right)\Bigg|_{x=x_0=p}\nonumber\\
&=&p^2-\frac{p^2}{2}=\frac{p^2}{2}.
\earr

\section{Implementing the Legendre Transform with the Faddeev-Popov Method}
We now want to show that another way of obtaining the Legendre transform is by applying the Faddeev-Popov trick to $F(x,p)$ as given by Eq. \eqref{FPalt}. To do this, we firstly rewrite Eq. \eqref{FPalt} for the case of ordinary functions (instead of functionals) as follows:
\begin{equation}\label{leg2}
\mathcal{I}=\int\;dx\;F(x,p)\Delta(p,x_0)\delta\Big(G(x,p)\Big),
\end{equation}
where $x_0$ is the point in which $G(x,p)=0$ with respect to $x$ and $\Delta(p,x_0)$ is a function to be yet determined. For the purpose of this note, we will assume that the function $G(x,p)$ vanishes only for a single given $x_0$ in the considered interval. We can expand the Dirac delta function in the integral with the usual expansion formula \cite{matBook}
\begin{equation}\label{dirac_fx}
\delta(w(x))=\frac{\delta(x-x_0)}{|w^{'} (x_0)|},
\end{equation}
where $x_0$ is the zero of the function $w(x)$ and $w'(x_{0})$ is the first derivative of $w(x)$ evaluated in $x=x_{0}$. Substituting this expression in eq. \eqref{leg2} brings to
\begin{equation}\label{FPx}
\mathcal{I}=\int\;dx\;F(x,p)\Delta(p,x_0)\frac{\delta(x-x_0)}{|G'(x_0,p)|},
\end{equation}
and it appear clear that we can choose $\Delta(x_0,p)$ in such a way that it compensates the extra term in the the denominator, namely
\beq
\Delta(p,x_0)=\left|\frac{\partial G(x,p)}{\partial x}\right|_{x=x_0}=|G'(x_0,p)|.
\eeq
Note that this definition is fully equivalent and consistent with the one given by Eq. \eqref{deltaD}. This brings to the following result:
\beq
\mathcal{I}=\int\;dx\;F(x,p)\delta(x-x_0)=F(x_0,p).
\eeq
Now, since $x_0$ is the value at which the function $G(x,p)=0$, the last term in the equality can be rewritten as $F(x,p)|_{G(x,p)=0}$, thus giving
\barr\label{legendre}
\mathcal{I}&=&\int\;dx\;F(x,p)\Delta(x_0,p)\delta\Big(G(x,p)\Big)dx\nonumber\\
&=&F(x_0,p)=F(x,p)|_{G(x,p)=0}\equiv g(p)
\earr
This is exactly the Legendre transform $g(p)$ of the initial function $f(x)$.  This result is worth a bit of discussion. To start with, let us consider again Eq. \eqref{leg1}. There, the Legendre transform is implemented by applying the constraint $G(x,p)=0$ to the function $F(x,p)$. In that case, the constraint is applied ``ad hoc" in order to produce the function $g(p)$. In Eq. \eqref{FPx}, instead, the same constraint $G(x,p)=0$ is applied to the function $F(x,p)$ in a more natural way by means of the Faddeev-Popov trick, by making the constraint the argument of the Dirac delta function. The Legendre transformation is now implemented by selecting (among all the possible values of $x$) only those values $x_0$ that make the argument of the Dirac delta in Eq. \eqref{FPx} vanish. The constraint, in this case, is therefore automatically applied, and the result is the same as in the case of the standard definition of Legendre transform as given by Eq.  \eqref{leg1}.

We now repeat the example of Sect. III by using the Faddeev-popov method given by Eq. \eqref{leg2} to calculate the Legendre transform of the function $f(x)=x^2/2$. As before, $F(x,p)=xp-x^2/2$ and $G(x,p)=p-x$, that vanish for $x_0=p$. Therefore we have that $\Delta(p,x_0)=|G'(p,p)|=1$ and
\beq
\delta\Big(G(x,p)\Big)=\delta(x-p).
\eeq
Substituting these results into Eq. \eqref{leg2} then gives:
\begin{equation}
\mathcal{I}=\int\;dx\;F(x,p)\delta(x-p)=F(p,p)=\frac{p^2}{2}\equiv g(p).
\end{equation}
This trivial, but instructive, example proves that our definition of Legendre transform via the integral in Eq. \eqref{legendre} is a valid and fully compatible definition of Legendre transform. Moreover, this allows us to give a more intuitive and clear explanation on how to apply the Faddeev-Popov method and what is its physical meaning, by providing an example of application in a more familiar context for the students, than the framework of quantum field theory where this method finds its natural application.

As a second, more physical example, let us consider the problem of calculating the Hamiltonian of an harmonic oscillator given its Lagrangian
\beq
\mathcal{L}_{ho}(q,\dot{q})=\frac{1}{2}m\dot{q}^2-\frac{1}{2}m\omega^2q^2,
\eeq
where $\dot{q}=dq/dt$, $m$ is the oscillator mass and $\omega$ is the characteristic oscillator frequency. As it is well known, the Hamiltonian is calculated by firstly introducting the canonically conjugate momentum $p=\partial\mathcal{L}_{ho}/\partial\dot{q}=m\dot{q}$ and then by performing a Legendre transform \cite{goldstein}:
\beq\label{hoLeg}
\mathcal{H}_{ho}(q,p)=p\dot{q}-\mathcal{L}_{ho}(q,\dot{q})=\frac{p^2}{2m}+\frac{1}{2}m\omega^2q^2.
\eeq
We now want to reproduce the same result by applying the Faddeev-Popov method. First of all, it is worth noticing that in this case the Legendre transform is made with respect to the variable $\dot{q}$, and that $q$ acts simply as a parameter in Eq. \eqref{hoLeg}. So said, we can build the analogue of the function $F(x,p)$ as
\beq
F(\dot{q},p;q)=p\dot{q}-\mathcal{L}_{ho}(q,\dot{q}).
\eeq
Notice that we used the semi-colon in the argument of $F(\dot{q},p;q)$ to indicate that here $q$ plays the role of a parameter, as the application of the Faddeev-Popov method leaves $q$ unchanged. The constraint $G(x,p)$ is now given by
\beq
G(\dot{q},p;q)=\frac{\partial F(\dot{q},p;q)}{\partial\dot{q}}=p-\frac{\partial\mathcal{L}_{ho}(q,\dot{q})}{\partial\dot{q}}=p-m\dot{q},
\eeq
and therefore $\Delta(\dot{q}_0,p;q)=|G'(\dot{q}_0,p;q)|=m$, where $\dot{q}_0=p/m$ is the point at which $G(\dot{q},p;q)$ vanishes with respect to $\dot{q}$. Notice that $G(\dot{q},p;q)=0$ corresponds to the usual definition of the canonical momentum in terms of derivative of the Lagrangian with respect to $\dot{q}$. This is the actual constraint that implements the Legendre transform.

In this case, Eq. \eqref{leg2} becomes
\barr
\mathcal{I}&=&\int\;d\dot{q}\Big[p\dot{q}-\mathcal{L}_{ho}(q,\dot{q})\Big]\;m\;\delta\Big(p-m\dot{q}\Big)\nonumber\\
&=&\int\;d\dot{q}\Big[p\dot{q}-\mathcal{L}_{ho}(q,\dot{q})\Big]\;\delta\Big(\dot{q}-\frac{p}{m}\Big)\nonumber\\
&=&p\left(\frac{p}{m}\right)-\mathcal{L}_{ho}\left(q,\frac{p}{m}\right)\nonumber\\
&=&\frac{p^2}{2m}+\frac{1}{2}m\omega^2q^2\equiv\mathcal{H}_{ho}.
\earr

\section{Conclusions}
In conclusion, we have discussed how the Faddeev-Popov method can be used to easily implement the Legendre transform of an arbitrary function $f(x)$. As an example, we considered the simple case of $f(x)=x^2/2$ as a direct verification of the validity of our correspondence. We have also presented, as a more physically meaningful example, how to use the Faddeev-Popov method to calculate the Hamiltonian of a classical harmonic oscillator starting from the knowledge of its Lagrangian function. This provides a novel point of view on the method itself, that only makes use of concepts familiar to the students such as the Legendre transform, without invoking any complicated theoretical framework as quantum field theory or gauge theory.



\begin{thebibliography}{99}
\bibitem{comment} Normally, the Lorentz gauge is considered. However, with no loss of generality we can assume any gauge condition here.
\bibitem{noteJ}In this case $J^{\mu}$ serves as ancillary variable that highly simplifies the calculations. See Ref. \cite{brown, ryder, peskin} for a more detailed discussion about the role of the current term $J^{\mu}$.
\end{thebibliography}
\end{document}